# Exploring Large-scale Gravitational Quantization without $\hbar$ in Planetary Systems, Galaxies, and the Universe

Howard G. Preston[1] & Franklin Potter[2,3]

**Abstract** We explore a theory of large-scale gravitational quantization, starting with GTR and the general relativistic Hamilton-Jacobi equation to create quantization conditions via a new scalar wave equation dependent upon two system parameters only, the total mass and the total vector angular momentum. Unlike quantum mechanics, Planck's constant is not required. Instead, a local invariant quantity proportional to the total angular momentum dictates the quantization conditions. In the Schwarzschild metric approximation, the theory predicts eigenstates with quantized energy per mass and quantized angular momentum per mass. Remarkably, a continuous orbit description is possible, in contrast to quantum mechanics. We find excellent agreement to the orbital spacings of the satellites of the Jovian planets and to the planet spacings in the Solar System. The existence of equilibrium radial distances for the eigenstates means that small radial accelerations exist to cause orbital evolution, which we apply to an artificial Earth satellite as an example. For galaxies we derive the baryonic Tully-Fisher relation and the MOND acceleration, so galaxy velocity curves are explained without requiring 'dark matter'. For the universe, in the interior metric approximation, we derive a new Hubble relation that accounts for the accelerated expansion with a matter density at about 5% of the critical matter/energy density. The remainder is attributed to large-scale quantization zero-point energy with a value of $\sim 1.1 \times 10^{-11}$ J m$^{-3}$ . However, none of the systems investigated can be shown to provide the definitive test of the theory, so we propose a possible laboratory test in which the two system parameters are well known.

[1] 15 Vista del Sol, Laguna Beach, CA 92651 USA
[2] Sciencegems.com, 8642 Marvale Drive, Huntington Beach, CA 92646 USA
[3] Please direct correspondence to: drpotter@citalumni.caltech.edu

## 1. Introduction

There are several "smoking guns" for large-scale gravitational quantization in familiar gravitationally-bound systems. For example, the angular momentum per mass for the planets of the Solar System is quantized [1] in terms of small integer ratios, and so is the angular momentum per mass for the major satellites of the Jovian planets. Our Galaxy (Milky Way) provides another example. The velocity of stars in the disk all have nearly the same rotational velocity of 220 ± 17 km s$^{-1}$ , as if they are all in the same quantized energy state per mass. Furthermore, the visible halo stars streaming [2] just beyond the Galaxy disk have the velocity 110 ± 15 km s$^{-1}$ , half the disk velocity, again providing a ratio of small integers and the possibility of a second quantized energy state per mass. Some of these small-integer ratio quantities may be explainable in terms of classical resonance phenomena, but not all.

We have two fundamental theories to work with. We know that the general theory of relativity (GTR) and quantum mechanics (QM) each describe natural phenomena with incredible accuracy. We know also that GTR and QM are incompatible in many ways, with the M-theory [3] of superstrings presently being the leading candidate for the unification of these two fundamental theories. However, a version of M-theory easily applicable to the possible gravitational quantization in planetary systems and galaxies is probably a decade away. Until then, a much simpler but reasonably comprehensive theory to help explain interesting gravitational phenomena inexplicable by GTR alone would be useful.

We explore one possible theory of large-scale gravitational quantization, a theory derived from the general relativistic Hamilton-Jacobi equation for which quantization conditions are created in a traditional mathematical way but with a new local invariant quantity instead of the universal quantity called Planck's constant $\hbar$. We show that numerous large-scale gravitationally-bound systems agree with its predicted quantization states of angular momentum per mass and energy per mass, including the satellites of the Jovian planets and the planets of the Solar System, as well as many galaxies [4, 5]. In fact, we derive the baryonic Tully-Fisher [6] relation (and therefore the MOND acceleration[7]) for galactic rotation velocities, i.e., a rotation velocity proportional to the fourth root of the baryonic mass of a galaxy. In addition, as an unforseen bonus, we derive the accelerated expansion behavior of the universe.

Before we discuss the theory and its many applications in detail, we insert here a brief preliminary summary of the major conceptual difference from QM. This new large-scale quantization theory dictates a quantization for each gravitationally-bound system that depends upon two physical quantities of the system only, its total mass M and its total vector angular momentum $H_\Sigma$. Instead of a universal constant such as Planck's constant $h$ and angular momentum quantization in terms of $\hbar$, the theory requires a different system scaling constant for each gravitationally-bound system

$$H = \frac{H_\Sigma}{M\,c}, \tag{1.1}$$

with the quantity $\mu$cH assuming the traditional role of $\hbar$, where c is the speed of light in vacuum and $\mu$ is the orbiting test particle mass. We show that large-scale quantization occurs as energy *per mass* and angular momentum *per mass* in terms of H.

Whether a consistent quantization theory without the universal constant $\hbar$ is feasible has never been discussed in the literature, as far as we know. Quantization with $\hbar$ is ingrained into our thinking and experience, so quantization with a local physical constant H instead is alien to our traditional understanding since 1900. Certainly, the ultimate test is whether the theory agrees well enough with the behavior of Nature to prove useful in understanding and predicting the relevant natural phenomena.

We show excellent agreement of the theory with numerous gravitationally bound systems, but our explorations with this theory have encountered its share of frustrations. Primarily, we expected to find and report a definitive test of this large-scale quantization theory for at least one large-scale system. However, no matter how successful has been the quantization fit to a physical system, none of our numerous application examples over distance scales up to $10^{21}$ meters and beyond can be considered as *the* definitive test of the theory because the total vector angular momentum of each system is not known well enough. As a consequence, we can show only that the systems comply with the proposed theoretical predictions. Therefore, we describe also a possible future laboratory experiment that might be able to develop into a definitive test.

## 2. New Scalar Gravitational Wave Equation

There are many ways in which to formulate a wave equation within the constraints of GTR. We have chosen to begin with the general relativistic Hamilton-Jacobi equation for a test particle of mass $\mu$ as given by Landau and Lifshitz [8]

$$g^{\alpha\beta}\,\frac{\partial S}{\partial x^\alpha}\,\frac{\partial S}{\partial x^\beta} - \mu^2\,c^2 = 0\,, \tag{2.2}$$

where $g^{\alpha\beta}$ is the metric of GTR and S is the action. Our choice originates from the historical fact that one can derive the Dirac equation of relativistic QM in the Lorentz metric limit of this equation, so the general method leading to a useful relativistic wave equation is well established. For the gravitational interaction, we build in the equivalence principle, converting S to a "field-like" quantity by dividing through by $\mu$c to obtain S' = S/$\mu$c. The above Hamilton-Jacobi equation then becomes a wave equation via the following transformation to eliminate the squared first derivative, i.e., by defining the wave function $\Psi$ (q, p, t) of position q, momentum p, and time t

$$\Psi = e^{\mathrm{i}S'/H}\,. \tag{2.3}$$

For comparison, the 2nd order Dirac equation can be derived using $\Psi = e^{\mathrm{i}S/\hbar}$ and the Lorentz metric in (2.2), then factoring with $\gamma$-matrices to obtain the 1st order Dirac equation. The standard Schrödinger equation follows by reducing to the non-relativistic limit. We will follow a different sequence to obtain a Schrödinger-like equation for the gravitational interaction in gravitationally-bound systems.

We continue with the transformation. Differentiating Ψ by the general coordinate $x^\alpha$ , then by $x^\beta$ , dropping the small second derivative of S' term because S' is assumed to be slowly varying, and multiplying by the metric produces

$$g^{\alpha\beta}\,\frac{\partial^2 \Psi}{\partial x^\alpha\,\partial x^\beta} + \frac{g^{\alpha\beta}}{H^2}\,\frac{\partial S'}{\partial x^\alpha}\,\frac{\partial S'}{\partial x^\beta}\,\Psi = 0\ . \tag{2.4}$$

Substituting the converted version of (2.2) with S' into the second term, the result is the new scalar 'gravitational wave equation' (GWE)

$$g^{\alpha\beta}\,\frac{\partial^2 \Psi}{\partial x^\alpha\,\partial x^\beta} + \frac{\Psi}{H^2} = 0 \tag{2.5}$$

The transformation (2.3) ensures that the GWE dictates the same acceleration for all orbiting bodies of different small masses in the same gravitational environment, in accordance with the GTR equivalence principle. From its solution in the appropriate metric, one can show also that the corresponding wave number k and frequency $\omega$ for the body in orbit are independent of its mass. In the limit of a free particle, one finds that $\Psi = C \exp[i\,(\mathbf{kx}\text{-}\omega t)]$ is a solution of the GWE in the Lorentz metric.

## 3. GWE in the Schwarzschild metric

In order to apply the GWE, we need to choose the appropriate GTR metric in the particular gravitationally-bound system and determine how the energy and angular momentum are quantized . Many systems of interest can be described very well in an approximation using the Schwarzschild metric.

$$c^2\,d\tau^2 = \left(1 - \frac{r_g}{r}\right) c^2\,dt^2 - \left(1 - \frac{r_g}{r}\right)^{-1} dr^2 - r^2\,d\theta^2 - r^2\,\text{Sin}^2\,\theta\,d\phi^2 \tag{3.6}$$

The GWE in the Schwarzschild metric becomes

$$\left(1 - \frac{r_g}{r}\right)^{-1} \frac{\partial^2 \Psi}{c^2\,\partial t^2} - \frac{1}{r^2}\,\frac{\partial}{\partial r}\left(r^2\left(1 - \frac{r_g}{r}\right)\frac{\partial \Psi}{\partial r}\right) - \frac{1}{r^2\,\text{Sin}\,\theta}\,\frac{\partial}{\partial \theta}\left(\text{Sin}\,\theta\,\frac{\partial \Psi}{\partial \theta}\right) - \frac{1}{r^2\,\text{Sin}^2\,\theta}\,\frac{\partial^2 \Psi}{\partial \phi^2} + \frac{\Psi}{H^2} = 0 \tag{3.7}$$

with Schwarzschild radius $r_g = 2GM/c^2$ . Note that first derivative terms will occur in the r and $\theta$ coordinates even though the GWE seems to dictate second derivative terms only in this diagonal metric. If the equation describes a classical mechanics environment, then the first derivative terms would be absent. Traditional quantum mechanics texts such as Kemble explain why equation (3.7) with first derivative terms is the correct equation when considering momentum and energy in the quantum mechanical environment.

The equation is separable and the appropriate substitution is the product wave function $\Psi = \Psi_t\,\Psi_r\,\Psi_\theta\,\Psi_\phi$ . Recall that S = $-E_0$ t + S(r) + S($\theta$) + L$\phi$. Separation of variables produces the coordinate equations, with the primes representing devision by $\mu$c:

$$\frac{d^2\,\Psi_t}{c^2\,d\,t^2} = \frac{-E_0'^2}{H^2\,c^2}\,\Psi_t \tag{3.8}$$

$$\frac{d^2\,\Psi_\phi}{d\,\phi^2} = \frac{-L'^2}{H^2}\,\Psi_\phi \tag{3.9}$$

$$\frac{1}{\text{Sin}\,\theta}\,\frac{d}{d\,\theta}\left(\text{Sin}\,\theta\,\frac{d\,\Psi_\theta}{d\,\theta}\right) - \frac{m^2}{\text{Sin}^2\,\theta}\,\Psi_\theta = -l\,(l+1)\,\Psi_\theta \tag{3.10}$$

$$\left(1-\frac{r_g}{r}\right)\frac{d^2\Psi_r}{dr^2}+\frac{(2-\frac{r_g}{r})}{r}\frac{d\Psi_r}{dr}+\left(\left(1-\frac{r_g}{r}\right)^{-1}\frac{E_0'^2}{H^2c^2}-\frac{1}{H^2}-\frac{l(l+1)}{r^2}\right)\Psi_r=0 \tag{3.11}$$

The relativistic energy $E_0'$ and the angular momentum $L'$ are associated with the particle in orbit in the gravitationally-bound system.

Requiring $\Psi_\phi$ to be a single-valued function dictates the angular momentum quantization condition

$$L'/H=m \tag{3.12}$$

with integer $m$ = 0, ±1, ±2, etc. Because L' = L/$\mu$c, the angular momentum *per mass* is quantized. The equation for $\Psi_\theta$ determines the azimuthal quantum number $l$ = 0, 1, 2, etc., with $|m| \le l$. When $|m| = l$ the maximum probability occurs at $\theta = \pi/2$, i.e., around the equatorial plane, where most of the orbiting mass for gravitationally-bound systems lies.

This radial equation has characteristics of a Klein-Gordon equation, so one cannot guarantee in general that the wave function $\Psi$ always has positive probability. For the electromagnetic interaction with its positive and negative electrical charges and currents, this failing of the Klein-Gordon equation is remedied by using the charge density and the current density instead of the probability density. For the gravitational interaction, even though mass and energy are equivalent, we are not sure that a negative mass density and negative mass current density are appropriate physical quantities. Therefore, we progress immediately to a Schrödinger-like equation that is first order in energy.

The radial equation has a singularity at r = $r_g$ that is transformed away by the standard substitution r ( r - $r_g$ ) = $r'^2$, where r' is a new radial coordinate. In all cases of interest, $r_g \ll$ r', typically r/r' < $10^{-8}$, so we choose to ignore terms proportional to $r_g/r'^2$, $r_g^2/r'^2$, and smaller. We make the traditional substitution for the relativistic energy $E_0 = \mu c^2$ + E, with E << $\mu c^2$, so the radial equation becomes

$$\frac{d^2\Psi_{r'}}{dr'^2}+\frac{2}{r'}\frac{d\Psi_{r'}}{dr'}+\frac{2}{H^2c^2}\left(\frac{E}{\mu}+\frac{r_g c^2}{2r'}-\frac{l(l+1)H^2c^2}{2r'^2}\right)\Psi_{r'}\approx 0, \tag{3.13}$$

which is Schrödinger-like and similar to the radial wave equation for the hydrogen atom. Instead of the electrostatic potential, this equation has the gravitational potential as the middle term in the large bracket, and $\mu$cH replaces $\hbar$. The solution for $\Psi_r$ is, after dropping the prime on the r for simplicity,

$$\Psi_r = A r^l \exp\left(\frac{-\sqrt{-2E/\mu}}{\mathrm{Hc}}r\right){}_1F_1\left(\frac{-r_g c}{2H\sqrt{-2E/\mu}}+l+1, 2l+2;\ 2\frac{\sqrt{-2E/\mu}}{\mathrm{Hc}}r\right) \tag{3.14}$$

with A the normalization constant. The quantity $\Psi\Psi^\dagger$ gives the probability density of the particle.

If the first parameter in the confluent hypergeometric function ${}_1F_1$ (g, h; z) is a negative integer or zero, then ${}_1F_1$ (g, h; z) reduces to a polynomial with a finite number of terms and the wave function does not diverge at infinity. Additionally, its second parameter must be a positive integer for the function to be single-valued. The second solution with U(g, h; z) does not yield a satisfactory wave function [9].

Setting the first parameter in ${}_1F_1$ (g, h; z) equal to $-n_r$, a negative integer or zero, and solving for the energy produces

$$E_n = -\mu c^2\frac{r_g^2}{8n^2H^2}, \tag{3.15}$$

on the order of $10^{-6}$ $\mu c^2$ or smaller for most cases of interest. The principal quantum number $n = n_r + l + 1$, always a positive integer. We see that the energy *per mass* is quantized. We see also that when $n$ = 1 there is a minimum energy state for the system consisting of a central mass and an orbiting body in the Schwarzschild metric, unlike the classical case.

For simplicity, we concentrate on circular or near-circular orbits only. When $l = n - 1$, the orbit is circular with a single radial peak in the probability distribution, and the confluent hypergeometric function reduces to a constant. We define a "gravitational Bohr radius" $r_0 = 2H^2/r_g$, which establishes the distance scale for the gravitationally-bound system because the peak in the wave function probability occurs at $n^2 r_0$ for each $n$. There is also an equilibrium radius of orbit $r_{eq} = n(n-1)\, r_0$ calculated from the negative gradient of the kinetic energy bracket in (3.13). And finally, one can define a 'gravitational de Broglie' wavelength $\lambda = 2\pi\ n\ 2H^2/r_g$, which is independent of the mass of the particle in orbit in agreement with the equivalence principle.

In the examples ahead, we determine the quantities H, $r_0$, and the energy eigenstates for many celestial systems and for the case of a $10^{-3}$ kg body orbiting a 10 kg body at one meter. We apply the large-scale quantization theory first to the simpler gravitationally-bound systems, the satellites of the Jovian planets and the planets of the Solar System, and then progress to more complicated systems. But before these applications are considered, we need to discuss the classical continuous orbit description in Newtonian mechanics and GTR in the context of this large-scale quantization theory.

## 4. Continuous Orbit Description

Conceptually, we would have great difficulty accepting any theory of celestial orbits that does not permit a classical continuous orbit description. If the successive positions of a massive orbiting body required the consideration of quantum probabilities, we would need to ask the famous question: Is the Moon really there when nobody looks? Experience tells us that GTR and Newtonian physics are extremely good at describing celestial orbits. Thus, a classical trajectory description with a continuous orbit is required. So how can we reconcile a classical trajectory in this large-scale quantization theory?

In the atomic realm in QM, the 'position' of an electron, for example, is associated with a large enough momentum uncertainty in order to obey the Heisenberg uncertainty principle $\Delta x\ \Delta p \geq \hbar/2$. The consequence is that observation by photons via the *electromagnetic* interaction in order to measure successive atomic positions of the electron will not describe a continuous classical trajectory.

Just as in QM, we can express an uncertainty principle in this large-scale quantization theory for the *gravitational* interaction as

$$\Delta x\, \Delta p \geqslant \mu\, c\ H/2\,. \tag{4.16}$$

Let's apply this uncertainty principle to the Earth-Moon system. We determine below that H ~ 20 meters for the Earth-Moon system, with H being much larger for most other celestial systems. Expressing $\Delta p = \mu\ \Delta v$, where $\Delta v$ is the velocity uncertainty, we have $\Delta x\ \Delta v \geqslant c\ H/2$, the right hand side being about $3 \times 10^9\ m^2\ s^{-1}$ for the Earth-Moon system, an enormous quantity. For example, if we know the Moon's position via graviton or gravitational wave scattering to within about $\Delta x \sim 1$ km, then its velocity uncertainty $\Delta v \geqslant 3 \times 10^6\ m\ s^{-1}$! This result will not allow a continuous orbit description.

Fortunately, we do not need to use the gravitational interaction to locate the position of the Moon. Practical photon observation of the position of a gravitationally-bound body in orbit results in a tiny uncertainty of its momentum compared to its total momentum value, i.e., $\hbar\omega/c \ll \mu v$, so successive determinations of its position are easily made.

Therefore, a classical description of the continuous particle trajectory is allowed in this large-scale quantization theory as long as gravitons are not required for the observation.

## 5. Planetary-like Orbital Systems

A test of a theory lies in its ability to describe the behavior of real physical systems. Our first applications of the GWE are to better understand the simpler gravitationally-bound systems having a large central mass and one or more small bodies in orbit and then, in later sections, progress to more complicated systems such as the Galaxy and the universe. In each application we will attempt to determine how well the description matches the bahavior and to establish where the greatest difficulties lie.

Planetary-like orbital systems, such as the Solar System or the satellites of Jupiter, are known to have radial orbital spacings that can be statistically fit by sets of small integers $n$ in several ways[10], including excellent fits [1] of orbital spacings proportional to $n^2$. As we state below for many systems, these excellent-fitting sets of values of $n$ for the orbiting bodies in a system are not unique. Even applying our large-scale quantization theory we cannot obtain unique statistical fits because the total angular momentum value $H_\Sigma$ is not known well enough, i.e., to within about 5% or so. However, we do show how these familiar systems comply with the theory in examples spanning an enormous range of distances from 1 meter to $10^{21}$ meters and more.

Recall that in all cases we consider circular orbits only. Our standard procedure will be to first apply the angular momentum quantization constraint (3.12), $m$ = L'/H, using the Newtonian $L/\mu = \sqrt{GMr}$, to determine the best set of integer values for the quantum number $m$ via a linear regression fit. We choose to define the best set as the integer set with the smallest $m$ values, preferably with a linear regression $R^2$ as high as possible, about 0.999. Remember that one can always go to a set of much larger integer values to obtain a better fit.

The total vector angular momentum of the system $H_\Sigma$ is determined from the angular momentum quantization fit. Then we use $l = |m|$ and $n = l + 1$ for circular orbits at the equatorial plane and determine the goodness of the corresponding distance vs [$n$ ($n$-1)] linear regression fit, i.e., proportional to the equilibrium radius for the circular orbit [11]. The two fits are not independent but provide a reliable method for finding the best values for $m$ and $n$.

An important concept here is that when the orbiting body is at its equilibrium orbital radius, then the large-scale quantization parameters agree with the classical Newtonian parameters of the orbit. Recall that in Newtonian physics and in GTR the orbits are in equilibrium at any radius. In this large-scale quantization theory, however, each eigenstate has its unique equilibrium radius of orbit. A body in circular orbit but not at the equilibrium radius, for example, will experience a small acceleration toward the equilibrium distance. We discuss this acceleration for an artificial Earth satellite in a later section.

*1. Satellites of the Jovian Planets.*

We assume that after billions of years the satellites of the Jovian planets are at their orbital equilibrium radii $r_{eq} = n$ ($n$-1) $r_0$. The four Jovian planet systems have practically all the mass and all the angular momentum in their central body, providing us with relatively simple systems to investigate. After the angular momentum fit to small integers and $R^2 >$ 0.999, the orbital distances were examined using the $n = l + 1$ values. The assignments for $n$ given in Table I for each of the Jovian systems all have the linear regression $R^2 > 0.999$. Not all eigenstates are occupied by known satellites, nor must they be occupied, for the history of the system plays an important role. That is, there may have been insufficient mass available for a satellite to form near the orbital state's equilibrium distance, or the satellite that initially formed may have been perturbed out of the state into a another state at a different equilibrium orbital radius. Notice that the planet radius also fits the scheme for these large planets [1].

From the angular momentum fit we determine the total angular momentum $H_\Sigma$ of the gravitationally-bound system. Unfortunately, the comparison to a known value with a small uncertainty cannot be made, for even though the moments of inertia I for the Jovian planets are known to within 0.5% by wobble measurements, differential rotation within each planet produces an uncertainty in its angular momentum I$\omega$ that could be more than 10%. Therefore, the best sets of integers listed in Table I are not unique, and other equivalent sets differing by one or more integers can have as good a fit as determined by their excellent values of $R^2 \sim 0.999$.

An interesting additional test of our quantization assignments for $n$ are the well-known classical resonances within these satellite systems, such as the 1:2:4 resonance of the massive Io, Europa, and Ganymede satellites of Jupiter. Substituing our expression for the equilibrium orbital radius R = $n^2 \; 2H^2 / r_g$ into Kepler's 3rd Law for any two satellites in resonance with orbital periods $T_1$ and $T_2$ produces $T_1 / T_2 = n_1^3 / n_2^3$. With Io at $n_1 = 9$, Europa at $n_2 = 11$ and Ganymede at $n_3 = 14$, the ratio $T_1 / T_2 = 0.548$ and $T_2 / T_3 = 0.485$, values close enough to 0.5 to accommodate the 1:2:4 resonance behavior.

*2. Planets of the Solar System.*

We assume that the planets of the Solar System have reached their orbital equilibrium radii $r_{eq} = n$ ($n$-1) $r_0$ over billions of years. The total mass of the Solar System is the mass of the sun, M = 2 x $10^{30}$ kg, but the angular momentum is mostly in the orbiting bodies. The nine planets have a total angular momentum of about 3.4 x $10^{43}$ kg-m$^2$ s$^{-1}$. Additional angular momentum exists in the Kuiper Belt at 30 - 100 AU and the Oort Cloud at 40,000 - 50, 000 AU. In fact, the Oort Cloud dominates the angular momentum with an expected value [12] of more than $10^{44}$ kg-m$^2$ s$^{-1}$.

We obtain excellent linear fits with $R^2 > 0.999$ to both the angular momentum constraint and to the planet distances with the quantization values given in Table II. From the slope of the angular momentum fit, we determine a Solar System total angular momentum $H_\Sigma$ = 1.86 x $10^{45}$ kg-m$^2$ s$^{-1}$, and H = 3.1 x $10^6$ m, requiring about 100 Earth masses in the Oort Cloud, a reasonable mass. Notice that even the eccentric orbit of Pluto is close enough to being circular to fit the scheme. Mercury begins the sequence with $n$ = 4, but a slightly different set of integers beginning with $n$ = 3 for Mercury has nearly as good a fit. Not all eigenstates are occupied by planets, so the history of the Solar System has been a factor. No planets in orbit closer than Mercury have been detected, although several states are available. With $r_0$ = 6.4 x $10^9$ m, and the solar radius at about 7 x $10^8$ m, the $n$ = 2 equilibrium orbital radius would be at 1.3 x $10^{10}$ m, well outside the Sun.

Again, we do not have a definitive test of the theory because there is some ambiguity in determining a unique set of quantum numbers, since alternative fits with slightly different sets of integers are possible. Also, the actual total angular momentum in the Oort Cloud is unknown.

*3. Extrasolar Planetary Systems.*

All the presently known extrasolar planetary systems with two or more planets, such as the Upsilon Andromedae system of at least three planets, can be fit with several different sets of values for $m$ and for $n$ because the total angular momentum of each system is unknown. We can, however, derive a surprising relationship among the masses of the orbiting bodies in any system from the angular momentum quantization condition (3.12). In the general case, summing over all the angular momenta of the bodies in the system, the spin of the central body $L_0$ plus the quantized orbital angular momentum for the ith body, produces the total angular momentum relation

$$H_\Sigma = L_0 + \frac{H_\Sigma}{M} \sum_i m_i \, \mu_i \, . \tag{5.17}$$

Expressing $L_0 = \beta \; H_\Sigma$, that is, $\beta$ gives the fraction of the angular momentum contributed by the central mass, this expression relates the total mass M to its satellite mass contributions by

$$M = \frac{1}{(1-\beta)} \sum_i m_i \, \mu_i \, . \tag{5.18}$$

For the Solar System, $\beta \sim 0$, because the angular momentum of the Sun is 1.0 x $10^{42}$ kg-m$^2$ s$^{-1}$, about one-thousandth the total. Using quantum values from Table II, the Solar System obeys this relationship only when the estimated mass and quantum numbers of the Kuijper Belt and Oort Cloud are included. For systems like the Jovian satellites, $\beta \sim 1$, so the relationship does not help us as much. But for systems similar to the Solar System, where $\beta \sim 0$, such as the extrasolar planetary systems, we can use the mass relationship to estimate how much more mass might be in the orbiting bodies or determine what the maximum quantum number set may be.

## 6. Binary Systems

Numerous gravitationally bound systems are binary systems, such as binary stars, the Earth-Moon system, the Pluto-Charon system, binary galaxies, etc. Using the Schwarzschild metric for these systems is a rough approximation, at best, especially when the value of the orbiting particle mass $\mu$ is a significant fraction of the total mass. The Schwarzschild metric factor (1 - $r_g$ / r) would probably become something like (1 - $r_g$ / r - $r'_g$ / r') as a first approximation, i.e., the $r_g$ / r is effectively the potential from one mass and the suggested expression would approximate the combined gravitational potential in the vicinity of two masses. The post-Newtonian formalism of GTR has a procedure for handling binary systems, but the equations are complicated.

We consider instead the reduced mass approach used in quantum mechanics[9] for a system of N particles. For the special case of N = 2 particles only, the procedure reduces to a simple result: the test particle mass $\mu$ is replaced by the reduced mass $\mu_R$ for the binary system, and the "effective potential" energy term $r_g c^2 / 2r$, i.e., the middle term in the kinetic energy bracket in (3.13), is replaced by the two-particle interaction energy $V_{12} = -GM_T/r$, where $M_T = M + \mu$. We therefore have the same equation and solutions as before in the Schwarzschild metric with just a notation change and the substitution of the appropriate values for the new $r_g$ and H.

*4. Two Small Masses.*

Consider two small masses gravitationally bound in orbit about each other, with M = 10 kg, $\mu = 10^{-3}$ kg, and r = 1 meter. Their reduced mass $\mu_R \sim \mu$, $M_T \sim$ M, and H ~ $\mu_R$ r v/ $M_T$ c, with v ~ $\sqrt{GM_T / r}$ . From the quantized angular momentum contraint L/$\mu_R$ = $m$cH, one obtains $m \approx M_T / \mu_R$ . Therefore, $m$ and $n \sim 10^4$ for this binary system, with H ~ 8.7 x $10^{-18}$ m, $r_g \sim 1.5$ x $10^{-26}$ m, and $r_0 \sim 10^{-8}$ m. At one meter separation, eigenstate equilibrium radii are spaced by about a millimeter.

*5. Earth-Moon system.*

We assume a system in equilibrium with the Moon at the equilibrium orbital radius r = 3.84 x $10^8$ m for a particular $n$ value. The reduced mass $\mu_R$ = 7.27 x $10^{22}$ kg is close to the Moon mass, and $M_T$ = 6.05 x $10^{24}$ kg, with H ~ 20 m. We get $m$ = 65 and $n$ = 66, with $r_0 \sim$ 8.7 x $10^4$ m, a radius well within the Earth average radius of 6.37 x $10^6$ m, so the $n$ = 2 through $n$ = 8 equilibrium radii are within the Earth radius.

*6. Pluto-Charon system.*

We assume again that an equilibrium orbital radius at 1.96 x x $10^7$ m has been reached. The reduced mass $\mu_R$ = 1.6 x $10^{21}$ kg, $M_T$ = 1.46 x $10^{22}$ kg, and H ~ 1.6 m. From the approximate quantized angular momentum constraint, $m$ = 9. The calculated $r_0$ = 2.6 x $10^5$ m.

*7. Binary Stars and Binary Galaxies.*

Consider a binary neutron star system where both masses are 2.8 x $10^{30}$ kg. From $m \approx M_T / \mu_R$ , one gets $m$ = 4 and therefore $n$ = 5. For two galaxies of nearly the same mass, one obtains the same result. The validity of the binary approximation in the calculation to estimate the angular momentum quantum number of these systems would need to be established before acceptance of these quantization values.

## 7. Galaxy disk rotation velocity

Although we have yet to find a definitive test of this large-scale quantization theory in familiar planetary-like gravitationally-bound systems, the theory's greatest potential value may be in helping us better understand the large-scale behavior of a galaxy. At present, galaxy dynamics and structure depend upon complicated plasma magnetodynamics and a halo of 'dark matter' that comprises about 90% of the mass [4, 5]. The galactic models are based upon classical physics without considering the possibility of large-scale quantization. Dark matter in sufficient quantities [5] to explain the non-Newtonian rotation velocity curve of the disk of a galaxy has been proposed but is yet to be detected directly, although additional stars and other masses have been detected in small quantities surrounding galaxies. So there remains the possibility that the 'dark matter' proposal is incorrect or only partially correct.

A successful model of many important aspects of galactic behavior without requiring 'dark matter' is Milgom's modified Newtonian dynamics (MOND) [6, 7] which applies whenever the radial acceleration is lower than the proposed MOND acceleration $a_0$ = -1.1 x $10^{-10}$ m $s^{-2}$, i.e., in practically any galaxy and galaxy cluster. Essentially, MOND replaces the Newtonian acceleration $g_N$ by

$$g = \sqrt{g_N \, a_0} \tag{7.19}$$

whenever $g_N < a_0$ . Of course, the content of MOND is much richer and one must read the literature to learn more.

An important but disturbing aspect of the MOND hypothesis modifying $g$ is that it has not been derived yet from first principles. However, using our large-scale quantization theory, we derive the MOND acceleration and the appropriate acceleration value using only the baryonic mass of a galaxy and its corresponding total angular momentum. The great advantage of our having agreement with the MOND model is that MOND fits the detailed velocity profiles of more than 100 galaxies of various types extremely well [13], so we can adopt its successes. Its greatest failure, not enough angular deflection of starlight by galactic gravitational lensing, may be remedied eventually by considering large-scale quantization.

First we apply the large-scale quantization theory to the Galaxy in order to show how the rotation velocity is predicted and to discuss some approximations in its application. The Galaxy (Milky Way) has a nucleus of stars rotating like a solid sphere of radius $R_{nuc} \sim 5 \times 10^{19}$ m and a disk of radius R ~ $5 \times 10^{20}$ m in which all stars have nearly the same tangential rotation velocity of 220 km $s^{-1}$. In order to utilize our previous GWE results without having to approximate a different metric within the disk itself, we begin with a test particle (i.e., a star) in orbit *outside* the visible extent of the galactic disk, i.e., at a distance slightly greater than its luminous baryonic mass radius R ~ $5 \times 10^{20}$ m. We assume that the Schwarzschild metric solutions are a reasonable approximation for this situation. Then the energy eigenvalue is given still by (3.15).

The application of the virial theorem leads directly to a tangential rotation velocity for the test particle

$$v = \frac{r_g\, c}{2\, n\, H} \,. \tag{7.20}$$

If we knew the mass and the angular momentum of the Galaxy well enough, we could evaluate the predicted velocity. However, the values of both the total baryonic mass and total angular momentum of the Galaxy are not known well enough. There is a wide range of acceptable mass values, $6 \times 10^{10} < M < 2.5 \times 10^{12}$ solar masses [14, 15], so we must estimate an upper and lower value for its total angular momentum. If we take baryonic matter at the minimum estimated mass of $6 \times 10^{10}$ solar masses and put this mass at the galactic radius R, we estimate an upper value for its angular momentum of $1.2 \times 10^{67}$ kg-$m^2$ $s^{-1}$ and a lower value of $8.0 \times 10^{65}$ kg-$m^2$ $s^{-1}$ when this minimum total mass orbits at the radius of the nucleus. If only baryonic matter contributes to the galactic mass, then its total angular momentum must be within the range just calculated.

We find that using M = $6 \times 10^{10}$ solar masses and $H_\Sigma \sim 4.4 \times 10^{66}$ kg-$m^2$ $s^{-1}$ accommodates the measured rotational velocity v ~ $2.2 \times 10^5$ m $s^{-1}$ with $n$ = 1. For the $n$ = 2 eigenstate, the predicted velocity is ~ $1.1 \times 10^5$ m $s^{-1}$ , in agreement with the recent values for the streaming stars just beyond the edge of the visible disk. From these results, we understand that the Galaxy is not a classical system but, instead, must be described by a large-scale quantization theory.

Therefore, we learn that the large-scale quantization theory can accommodate the measured galaxy rotation velocity value at the lowest possible baryonic mass value. At the opposite extreme lies the 'dark matter' model which requires a total Galaxy mass of at least $7.2 \times 10^{11}$ solar masses, with about 10% being the luminous baryonic matter. One reasonable conclusion is that the behavior of a galaxy can be understood without requiring 'dark matter'.

If we were to develop a more detailed model of the Galaxy, we would abandon the Schwarzschild metric approximation and would approximate the galactic potential in the disk region by a nuclear shell potential of nuclear physics that averages the contribution from all the particles. One can show that at most two or three bound states survive in the shell-model potential well, the states with quantum numbers $n$ = 1, $l$ =0 and $l$ = 1 and possibly $n$ = 2, $l$ = 0. Then one would interpret the $n$ = 1, $l$ =0 state as corresponding to the galactic nucleus because having $l$ =0 would allow mass to accumulate at small radius into a sphere. The $n$ = 1, $l$ = 1 state would correspond to the disk state with matching to the Schwarzschild state at its edge. The $n$ = 2 state would contain the streaming stars beyond the disk. The same approach would be applied to a cluster of galaxies to determine its quantization states. We leave these developments for future research.

Conceptually, however, the most important practical application of the large-scale quantization theory is to derive the baryonic Tully-Fisher relation showing that the rotation velocity $v$ is proportional to the fourth root of the galaxy's mass, which has been determined to hold true for practically all galaxies [6]. The MOND model predicts this relation. Then we can derive the MOND acceleration expression.

As before, our major difficulty is the evaluation of H = $H_\Sigma$ /Mc. However, we can work with this limitation by assuming that all the baryonic mass M is uniformly distributed in a narrow ring of radius R so that $H_\Sigma = \alpha$MRv, where $\alpha$ is a parameter to be adjusted to obtain the actual numerical value. Then, in the velocity expression (7.20) we use M = $\pi h R^2 \rho_0$ for a narrow ring of thickness h and uniform density $\rho_0$ , substitute for R, and immediately derive

$$v = M^{1/4} \sqrt{\frac{G}{n\alpha} \sqrt{\pi h \rho_0}} \quad . \tag{7.21}$$

This expression is our version of the famous baryonic Tully-Fisher relation relating the fourth root of the galaxy mass to the rotation velocity, with the square root acting as a proportionality constant because the quantity inside varies only slightly.

What can we learn from our derived baryonic Tully-Fisher relation? Using the minimum mass value above and reasonable values for the thickness and the average density, we require $\alpha \sim 0.3$, telling us that placing all the mass in a rotating ring at R produces too much angular momentum by a factor of about three. One could do a more sophisticated model of the Galaxy mass and angular momentum distributions in order to exploit this fundamental relationship more thoroughly.

We now derive the MOND acceleration using the velocity expression (7.20) also. The effective gravitational acceleration g in (7.19) equals $v^2$ /r for equilibrium in a circular orbit. Substitution for the velocity squared produces

$$g = \frac{r_g^2\, c^2}{4\, n^2\, H^2\, r} \; . \tag{7.22}$$

Place the square of the right-hand side under a square root and factor out the Newtonian acceleration to get

$$g = \sqrt{\frac{GM}{r^2} \left( \frac{G^3\, M^7}{n^4\, H_\Sigma^4} \right)}, \tag{7.23}$$

from which we identify the MOND acceleration

$$a_0 = \frac{G^3\, M^7}{n^4\, H_\Sigma^4} \, . \tag{7.24}$$

For the Galaxy, using the values listed above with $n = 1$, we calculate an $a_0 = -2.8 \times 10^{-10}$ m s$^{-2}$, about two and a half times Milgrom's average value $a_0 = -1.1 \times 10^{-10}$ m s$^{-2}$ that was determined by averaging over many galaxies. A small adjustment in our estimate of M or $H_\Sigma$ would accommodate this difference. Our results suggest that $a_0$ would be slightly different for each galaxy instead of being taken as a universal value.

We have used the Schwarzschild metric to approximate the behavior of a test mass in orbit about the Galaxy and we have detemined an expression for its velocity that correctly predicts the value in the $n = 1$ eigenstate. From this velocity expression we derived the baryonic Tully-Fisher relation and the MOND acceleration for the first time by considering the baryonic mass and its total angular momentum only. The overall result is that galaxies exhibit large-scale quantization and do not require 'dark matter'.

## 8. Expansion of the Universe

One wonders whether our large-scale quantization theory that is successful in explaining the behavior of many gravitationally-bound systems can be applied to the universe itself. So we consider the universe in the GTR interior metric approximation, which is directly related to the general Robertson-Walker type of metric. We show that a new Hubble relation is predicted that agrees with the accelerated expansion of the universe as determined by the Supernova Type 1a experiments.

The general form of the interior metric for a spherical mass distribution is

$$c^2\, d\tau^2 = A(r)\, c^2\, dt^2 - B(r)\, dr^2 - r^2\, d\theta^2 - r^2\, \mathrm{Sin}^2\, \theta\, d\phi^2 \tag{8.25}$$

where A and B are to be determined. For *any* location, i.e., all observers, inside a dust-filled, constant density universe with no pressure,

$$B(r) = \frac{1}{1 - 8\pi G \rho r^2 / 3c^2}, \tag{8.26}$$

and, for a continuous constant density $\rho$,

$$A(r) = \alpha (1 - 8\pi G \rho r^2 / 3c^2). \tag{8.27}$$

Note that a constant $\alpha$ with a value between 1/4 and 1 multiplies the righthand side of (8.27) and later would add a constant term to the radial velocity we calculate below. We will drop this constant multiplier $\alpha$ now in order to simplify the expressions, for the resulting constant term would be subtracted away later by the appropriate integration constant value and therefore can be ignored.

We define $\mathrm{k} = 8\pi G \rho / 3c^2$. The GWE in this interior metric is

$$(1 - \mathrm{kr}^2)^{-1} \frac{\partial^2 \Psi}{c^2 \partial t^2} - \frac{1}{r^2} \frac{\partial}{\partial r}\left(r^2 (1 - \mathrm{kr}^2) \frac{\partial \Psi}{\partial r}\right) - \frac{1}{r^2 \operatorname{Sin}\theta} \frac{\partial}{\partial \theta}\left(\operatorname{Sin}\theta \frac{\partial \Psi}{\partial \theta}\right) - \frac{1}{r^2 \operatorname{Sin}^2\theta} \frac{\partial^2 \Psi}{\partial \phi^2} + \frac{\Psi}{H^2} = 0 \tag{8.28}$$

After working out the parts for the other three coordinates, the full radial part of the GWE is

$$(1 - \mathrm{kr}^2) \frac{d^2 \Psi_r}{d r^2} + \frac{(2 - 4\mathrm{kr}^2)}{r} \frac{d \Psi_r}{d r} + \frac{2}{H^2 c^2}\left((1 - \mathrm{kr}^2)^{-1} \frac{E_0'^2}{2} - \frac{c^2}{2} - \frac{l(l+1) H^2 c^2}{2 r^2}\right) \Psi_r = 0, \tag{8.29}$$

and using $E_0 = \mu c^2 + \mathrm{E}$, with $\mathrm{E} << \mu c^2$,

$$\frac{d^2 \Psi_r}{d r^2} + \frac{2}{r} \frac{d \Psi_r}{d r} + \frac{2}{H^2 c^2}\left(\frac{E}{\mu} + \frac{\mathrm{kr}^2 c^2}{2(1 - \mathrm{kr}^2)^2} - \frac{l(l+1) H^2 c^2}{2 r^2 (1 - \mathrm{kr}^2)}\right) \Psi_r \approx 0. \tag{8.30}$$

We identify an 'effective potential' in the universe for each observer

$$V_{\mathrm{eff}} \approx -\frac{\mathrm{kr}^2 c^2}{2(1 - \mathrm{kr}^2)^2} + \frac{l(l+1) H^2 c^2}{2 r^2 (1 - \mathrm{kr}^2)}, \tag{8.31}$$

for which we expect the angular momentum term to be zero or nearly zero. The negative gradient of the first term in $V_{\mathrm{eff}}$ is the average radial acceleration

$$< \ddot{r} > = k c^2 r \frac{(1 + k r^2)}{(1 - k r^2)^3}, \tag{8.32}$$

from which we derive a new Hubble relation

$$< \dot{r} > = r \frac{c \sqrt{k}}{1 - k r^2}, \tag{8.33}$$

with the value of the integration constant taken so that the Hubble velocity at any observer's origin is zero. Of course, the maximum velocity is limited to the speed of light. Notice that an alternative solution could have been obtained relating the radial velocity to an inflationary exponential function of the time coordinate.

For distances up to about one billion light-years, when $kr^2 << 1$, we recover the standard Hubble relation, and if we take the critical density $\rho = 8 \times 10^{-27}$ kg m$^{-3}$, then k ~ $5 \times 10^{-53}$ m$^{-2}$ and the Hubble constant h ~ $2 \times 10^{-18}$ s$^{-1}$, about 62 km per second per megaparsec, an acceptable value. At the further distance of $5 \times 10^9$ lt-yrs, $kr^2 \sim 0.11$, producing a 12% increase, consistent with the increase measured in the Supernova Type 1a experiments.

We should predict therefore that the matter/energy density of the universe is actually the critical density in order to have the correct value for the Hubble constant. However, there is a problem. Present measurements of the matter/energy density with known particles falls far short of the critical density. Several concepts that have been suggested in order to explain this discrepancy include a non-zero cosmological constant, 'dark energy', and quantum mechanical zero-point energy. The standard QM zero-point energy density calculations are known to be at least 30 orders of magnitude too big.

Perhaps the zero-point energy density associated with our large-scale gravitational quantization will be closer. There are several ways to estimate the zero-point energy density of the vacuum, including the application of the uncertainty principle. We can make an estimate of the zero-point energy in this large-scale quantization theory using the effective potential, that is,

$$E_{\text{zero}} \sim \rho\, V_{\text{eff}} \approx -\frac{\rho\, \text{kr}^2\, c^2}{2\,(1-\text{kr}^2)^2}\,, \tag{8.34}$$

where $\rho$ is the present known matter density, approximately 5% of the critical density. Substituting this small 5% value for $\rho$ into the expression, including this value in k also, and setting r ~ $10^{26}$ m, i.e., about 10 billion lt-yrs for the radius of the universe, produces a total vacuum energy density

$$E_{\text{zero}} \sim 1.1 \times 10^{-11}\ \text{J m}^{-3}\,, \tag{8.35}$$

very nearly the ~ $2 \times 10^{-11}$ J $\text{m}^{-3}$ needed for the Hubble relation to agree with radial velocities.

In conclusion, we argue that the total matter/energy density in the universe is at the critical value, with about a 5% to 10% contribution from known matter and the rest from the large-scale quantization zero-point energy. There is no need for exotic particles or a non-zero cosmological constant to explain the critical density.

## 9. Artificial Earth Satellite

We return to the Schwarzschild metric solutions of the GWE to consider the behavior of an Earth satellite in an arbitrary orbit and its slow evolution to an equilibrium distance. In the Schwarzschild metric in GTR, every circular orbit is an equilibrium orbit, ignoring the perihelion precession, Poynting-Robertson effect, Libby-Wood effect, etc. In our large-scale quantization theory, however, a body in an arbitrary circular orbit will not have energy per mass and angular momentum per mass values corresponding to a single eigenstate of the system. Only when the orbiting body is at the radial equilibrium distance $r_{\text{eq}}$ will the Newtonian energy and angular momentum agree with the quantized eigenvalues.

In the large-scale quantization theory a body in an arbitrary orbit is in a linear superposition of eigenstates and will experience a small acceleration toward the $r_{\text{eq}}$ of one eigenstate. The "effective potential" in (3.13) determines the classical radial turning points of the orbit for each possible eigenstate, that is, the radial distance range in which the kinetic energy remains real. One first determines which few eigenstates are allowed and then calculates the orbital evolution. We consider an artificial Earth satellite and other examples below.

A detailed comparison of Newtonian gravitation and the large-scale quantization theory can be made for a single test particle of mass $\mu$ orbiting about a spinless massive central body in the Schwarzschild metric approximation. The total energies per mass are, respectively,

$$\frac{E}{\mu} = -\frac{\text{GM}}{2\,r} + \frac{L^2}{4\,\mu^2\,r^2}\,. \tag{9.36}$$

$$\left\langle \frac{E}{\mu} \right\rangle = -\left\langle \frac{\text{GM}}{2\,r} \right\rangle + \left\langle \frac{l\,(l+1)\,H^2\,c^2}{4\,r^2} \right\rangle, \tag{9.37}$$

where the < E/$\mu$> is the expected value of the operator, as in QM. The "effective potentials" are, respectively,

$$V(r) = -\frac{\text{GM}}{r} + \frac{L^2}{2\,\mu^2\,r^2} \tag{9.38}$$

$$\langle V(r) \rangle = -\left\langle \frac{\text{GM}}{r} \right\rangle + \left\langle \frac{l\,(l+1)\,H^2\,c^2}{2\,r^2} \right\rangle. \tag{9.39}$$

From the potentials we use the virial theorem and calculate the radial accelerations via the negative gradient to obtain

$$\ddot{r} = -\frac{\mathrm{GM}}{r^2} + \frac{L^2}{\mu^2\, r^3} \tag{9.40}$$

$$\langle \ddot{r} \rangle = -\left\langle \frac{\mathrm{GM}}{r^2} \right\rangle + \left\langle \frac{l\,(l+1)\,H^2\, c^2}{r^3} \right\rangle. \tag{9.41}$$

In Newtonian physics the angular momentum $\mathrm{L} = \mu\,\sqrt{\mathrm{GMr}}$ and $\dot{\phi} = \mathrm{L}/\mu r^2$ for a circular orbit, so $\ddot{r} = 0$ and the radial acceleration per mass ($\ddot{r}$ - r $\dot{\phi}^2$ ) equals -GM/$r^2$ .

In the large-scale quantization theory a continuous orbit description is allowed, so we set up the integrals for the operators and insert a Dirac delta function for the particular radial distance. One finds that $\ddot{r} \neq 0$ for any r value other than for $r_{\mathrm{eq}}$ . At other values of radial distance there will be an additional acceleration toward the equilibrium radius of the eigenstate. Energy and angular momentum are exchanged with the rest of the system to maintain conservation of energy and angular momentum.

Consider an artificial Earth satellite. We would like to know what happens to this satellite after being placed in a circular orbit at a chosen radial distance of 100 km above the Earth's surface, that is, at 6.47 x $10^6$ m compared to an Earth radius of 6.37 x $10^6$ m. The artificial satellite will have a classical Newtonian energy per mass $E_{\mathrm{class}}$ = -3.08 x $10^7$ J/kg and an angular momentum pre mass $L_{\mathrm{class}}$ = 5.08 x $10^{10}$ kg-$m^2\, s^{-1}$ /kg, an energy value and angular momentum value that do not equal eigenvalues for any one of the possible eigenstates of the Earth-Moon system. We calculate that the satellite can be in the $n$ = 8, 9, or 10 eigenstates only, otherwise its radial position for a circular orbit is outside the range of the classical turning points $R_{\mathrm{min}}$ and $R_{\mathrm{max}}$ and its kinetic energy would be imaginary.

We determine the coefficients a, b, and c for the linear combination of the three QCM states,

$$\psi = a\,\psi_8 + b\,\psi_9 + c\,\psi_{10}, \tag{9.42}$$

$$E_{\mathrm{class}} = E_8 \; |\,a\,|^2 + E_9 \; |\,b\,|^2 + E_{10} \; |\; c \;|^2, \tag{9.43}$$

$$L_{\mathrm{class}} = L_8 \; |\,a\,|^2 + L_9 \; |\,b\,|^2 + L_{10} \; |\; c \;|^2, \tag{9.44}$$

$$|\,a\,|^2 + |\,b\,|^2 + |\; c \;|^2 = 1. \tag{9.45}$$

This linear combination applied at 6.47 x $10^6$ m produces a radially inward $\ddot{r} \simeq$ -2.22 x $10^{-7}$ m $s^{-2}$ , a very small correction to the Newtonian value of zero, leading to a slow evolution toward $r_{\mathrm{eq}}$ at about 6.29 x $10^6$ m and subsequent collision with Earth. The actual value of $\ddot{r}$ changes with radial distance in a nonlinear way.

Perhaps this small acceleration can be detected. The Earth satellite starting at 100 km above the surface is subject to many perturbations of its motion that are larger that this correction factor, including atmospheric compression effects. Periodic perturbations by mountain ranges, etc., can be accounted for and subtracted, so the Earth satellite may be an instrument for a definitive test but the details need to be worked out.

We have applied this calculation procedure to other systems and have found that a very slow evolution to the equilibrium orbital radius occurs in many other interesting physical cases. For example, if we assume a present radial position R = 1.496 x $10^{11}$ m of the Earth in a circular orbit, then the Earth would move to the $n$ = 6 equilibrium at $r_{\mathrm{eq}}$ = 1.452 x $10^{11}$ m with a present radial acceleration less than -1.1 x $10^{-8}$ m $s^{-2}$, taking more than 100,000 years. The acceleration value decreases as $r_{\mathrm{eq}}$ is approached. Even so, the Earth will overshoot slightly and experience oscillations about $r_{\mathrm{eq}}$ .

We have found also that most radial accelerations for bigger systems, such as the Galaxy, are often smaller, on the order of $10^{-10}$ m $s^{-2}$ for stars in the disc. As a result, the radial movement of stars in the disk of the Galaxy is almost nonexistent on the large scale because local effects produced by nearby stars or star clusters introduce larger radial accelerations.

## 10. Possible Laboratory Experiment

A laboratory experiment to test the theory requires measurable gravitational forces greater than about $10^{-12}$ N and a 'gravitational Bohr radius' $r_0$ of about one meter or less so that equilibrium radii for $n > 1$ exist within room dimensions. Suppose we have a $10^{-2}$ kg mass 'in orbit' about a spinning $10^3$ kg central mass at a 0.5 m orbit radius. Since $r_g = 1.5 \times 10^{-24}$ m, in order to have an $r_0$ of 0.10 m, say, we need an H = 2.7 x $10^{-13}$ meters and $H_\Sigma = 8.2 \times 10^{-2}$ kg-$m^2$ $s^{-1}$. A $10^3$ kg spinning iron sphere would need to spin at about 5.1 x $10^{-3}$ radians per second, or about 3 revs/hr. The classical turning points for the total kinetic energy dictate that only the $n$ = 2 and $n$ = 3 eigenstates can participate. For $n$ = 2, the GWE radial acceleration at 0.5 m is about -7.2 x $10^{-9}$ m $s^{-2}$ compared to the Newtonian radial acceleration of -2.69 x $10^{-7}$ m $s^{-2}$, a measurable effect. One would alter the spin rate of the central sphere to emphasize the sign change of the GWE accelerations on different sides of the equilibrium radii. A torsion balance placed near the spinning sphere might be a reasonable configuration, but there is the question of whether such a system can be considered gravitationally bound and in circular orbit. We are examining other possible definitive tests, including an Earth satellite as mentioned above.

## 11. Final Comments

We have reported our explorations into a new theory of large-scale quantization in gravitationally-bound systems based upon a new scalar gravitational wave equation requiring two system parameters only, the total mass and the total vector angular momentum. In the approximation with the Schwarzschild metric, the theory predicts states with quantized energy per mass and quantized angular momentum per mass. Unlike quantum mechanics, Planck's constant $h$ is not used and a continuous orbit description is possible. We find good agreement of theory predictions for the orbital spacings of the satellites of the Jovian planets and of the planets of the Solar System as well as reasonable approximate results for some binary systems.

We applied the theory to galaxies, for which we derived the baryonic Tully-Fisher relation connecting the rotational velocity to the fourth root of the baryonic mass. We also derived the MOND acceleration using the baryonic mass only, so the tangential velocity curve of the disk stars in any galaxy is explained without requiring 'dark matter'. The streaming halo stars just outside the visible disk were put into the $n$ = 2 eigenstate with one-half the disk rotation velocity, in agreement with recent data. Consequently, galaxies are quantized systems and are not Newtonian!

Using the interior metric as an approximation to a uniform density universe, we derived a new Hubble relation that explains the accelerated expansion measured by the Supernova Type 1a experiments and agrees with the standard Hubble constant for nearby galaxies as long as the matter/energy density is taken to be the critical density. The 'missing' matter/energy density was shown to correspond to the large-scale gravitational quantization zero-point energy instead of the standard quantum mechanical zero-point energy, so one does not need a non-zero cosmological constant.

None of these systems described in the Schwarzschild metric approximation provides a definitive test of the theory because each has a total angular momentum that is not known well enough. We have therefore outlined a possible laboratory test of the theory in which the total angular momentum is well known, using a method that would capitalize on passing through several equilibrium radii for the quantization states of a small mass (i.e., a sensitive torsion balance) near a large spinning mass. A precision test using an Earth satellite may be possible also.

We conclude that the large-scale gravitational quantization theory is a useful simple theory because many known gravitational problems are resolved. Certainly, we have shown how GTR can be extended to include quantization effects in some practical way in order to progress in our understanding of astrophysical phenomena. Hopefully, we have begun the exploration in the right direction.

We would like to thank colleagues in the Department of Physics and Astronomy at the University of California at Irvine who have given us useful suggestions and helpful criticism during many discussions. We would also like to thank Professor Joseph Weber (now deceased) of the University of Maryland for penetrating questions regarding the behavior of an artificial Earth satellite under the quantization conditions.

TABLE I. Jovian planet satellite assignments by the GWE.

| Name | $n$ | Name | $n$ |
|---|---|---|---|
| **Jupiter** | | **Uranus** | |
| radius | 4 | radius | 5 |
| Amalthea | 6 | Belinda | 8 |
| Io | 9 | Ariel | 12 |
| Europa | 11 | Umbrial | 14 |
| Ganymede | 14 | Titania | 18 |
| Callisto | 18 | Oberon | 21 |
| **Saturn** | | **Neptune** | |
| radius | 4 | radius | 6 |
| Mimas | 6 | Naiad | 8 |
| Enceladus | 7 | Galatea | 9 |
| Tethys | 8 | Larissa | 10 |
| Dione | 9 | Proteus | 12 |
| Rhea | 10 | Triton | 20 |
| Titan | 15 | □ | □ |
| Iapetus | 25 | | |

TABLE II. Planet assignments in the Solar System by the GWE

| Name | $n$ |
|---|---|
| Mercury | 4 |
| Venus | 5 |
| Earth | 6 |
| Mars | 7 |
| Jupiter | 12 |
| Saturn | 16 |
| Uranus | 22 |
| Neptune | 27 |
| Pluto | 31 |